\documentstyle[12pt,fleqn]{article}
\def\la{\langle}
\def\ra{\rangle}
\def\beeq{\begin{equation}}
\def\eneq{\end{equation}}
\def\beeqa{\begin{eqnarray}}
\def\eneqa{\end{eqnarray}}

\addtocounter{section}{-1}
\begin{document}

\begin{center}

\vspace{2cm}

{\large {\bf {Long-range excitons in conjugated polymers\\
with ring torsions: poly({\mbox{\boldmath $para$}}-phenylene) 
and polyaniline\\
} } }

\vspace{1cm}
(Running head: {\sl Long-range excitons in conjugated polymers})
\vspace{1cm}

{\rm Kikuo Harigaya\footnote[1]{E-mail address: 
\verb+harigaya@etl.go.jp+; URL: 
\verb+http://www.etl.go.jp/+\~{}\verb+harigaya/+}}

\vspace{1cm}

{\sl Physical Science Division,
Electrotechnical Laboratory,\\ 
Umezono 1-1-4, Tsukuba 305-8568, Japan}

\vspace{1cm}

(Received~~~~~~~~~~~~~~~~~~~~~~~~~~~~~~~~~~~)
\end{center}

\vspace{1cm}

\noindent
{\bf Abstract}\\
Ring torsion effects on optical excitation properties in
poly({\sl para}-phenylene) (PPP) and polyaniline (PAN) 
are investigated by extending the Shimoi-Abe model
[Synth. Met. {\bf 78}, 219 (1996)].  The model is solved by 
the intermediate exciton formalism.  Long-range excitons 
are characterized, and the long-range component of the 
oscillator strengths is calculated.  We find that ring 
torsions affect the long-range excitons in PAN more easily 
than in PPP, due to the larger torsion angle of PAN and 
the large number of bonds whose hopping integrals are 
modulated by torsions.  Next, ring torsional disorder effects
simulated by the Gaussian distribution function are analyzed.
The long-range component of the total oscillator strengths
after sample average is nearly independent of the disorder
strength in the PPP case, while that of the PAN decreases 
easily as the disorder becomes stronger.

\vspace{1cm}

\pagebreak

\section{Introduction}

Recently, we have been studying structures of photoexcited states 
in electroluminescent conjugated polymers: poly({\sl para}-phenylene) 
(PPP), poly({\sl para}-\-phen\-yl\-ene\-vinylene) (PPV), 
poly({\sl para}-phenylenedivinylene) (PPD), and so on [1-3].
We have introduced the viewpoint of long-range excitons in
order to characterize photoexcited states where an excited
electron-hole pair is separated over a single monomer of the 
polymer.  We have shown [1,2] that a long-range exciton feature 
starts at the energy in the higher energy side of the lowest 
feature of the optical absorption of PPV.  The presence of 
the photoexcited states with large exciton radius is essential 
in mechanisms of the strong photocurrents observed in this 
polymer.  In Ref. [3], we have compared properties of excitons 
in PPV-related polymers.  The  oscillator strengths of the 
long-range excitons in PPP are smaller than in PPV, and those 
of PPD are larger than in PPV.  Such relative variation is 
due to the difference of the number of vinylene bonds.

It is known that the PPV chain is nearly planer and the phenyl
rings are not distorted each other, as observed in X-ray 
analysis [4].  However, ring torsions, where even number phenyl 
rings are rotated in the right direction around the polymer axis 
and odd number phenyl rings are rotated in the left direction,
have been observed in PPP [5].  The polymer structure of PPP is 
shown schematically in Fig. 1 (a).  The ring torsions originate 
from the steric repulsion between phenyl rings, because the 
distance between the neighboring rings in PPP is smaller than 
that of PPV.  In a simple tight binding model, the ring torsion 
modulates the nearest neighbor hopping integral $t$ as
$t {\rm cos}\Psi$, where $\Psi$ is the torsion angle.  If the 
angle $\Psi$ is sufficiently large, the motion of electrons 
between phenyl rings would be hindered and also the exciton 
radius of photoexcited states would become shorter.  The 
contribution from long-range excitons will be smaller from 
that of the calculations [3] where the planer structure of 
PPP has been assumed.  The actual torsion angle about 23$^\circ$ [5]
may influence the results of the previous calculations.
The first purpose of this paper is to examine ring torsion
effects on optical excitations of PPP.  We will show that
the magnitude of the torsion $\Psi \sim 23^\circ$ does not
change the component of the long-range excitons so much.

The another example of polymers where ring torsions are present
is polyaniline (PAN) (leucoemeraldine base).  The schematic 
structure is displayed in Fig. 1 (b).  The phenyl rings and NH 
units are arrayed alternatively in the chain direction.  The 
magnitude of the torsion of phenyl rings in PAN is about 
56$^\circ$ [6].  This is larger than that of PPP, and thus the 
photoexcited states in PAN will be influenced more strongly by the 
torsion.  In the last half of this paper, we shall look at this 
problem.  We will show that the long-range component of the 
oscillator strengths at $\Psi = 56^\circ$ is about half 
of the magnitude of the system without ring torsions. 
Long-range excitons in PAN are hindered by ring torsions
more easily than those in PPP.  This is due to the fact that
the torsion angle is larger in PAN, and also that two bonds, 
whose hopping integrals are modulated as $t {\rm cos}\Psi$,
are present between neighboring phenyl rings in PAN while one
such bond is present between phenyls in PPP.

Furthermore, it is possible that torsion angles in these 
polymers are not uniform over all the systems.  The torsion 
angles might fluctuate spatially owing to some perturbation
effects by thermal origins or by external potentials. 
In this paper, such disorder effects are simulated by the 
Gaussian distribution functions.  The disorder strength is
changed extensively, and the average long-range component
is calculated for PPP and PAN.  We show that the long-range 
after sample average is nearly independent of the disorder
strength in the PPP case, while that of the PAN decreases 
apparently as the disorder becomes stronger.

This paper is organized as follows.  In the next section, 
our model is described and the characterization method of
long-range excitons is explained.  In \S 3, ring torsion 
effects on photoexcited states in PPP are reported.  The 
results of PAN are given in \S 4, and the paper is concluded 
with a summary in \S 5.

\section{Formalism}

We consider optical excitations in PPP and PAN, by extending
the Shimoi-Abe model [7] with electron-phonon and electron-electron 
interactions.  The model is shown below:
\beeqa
H &=& H_{\rm pol} + H_{\rm int}, \\
H_{\rm pol} &=& \sum_{i,\sigma} E_i c_{i,\sigma}^\dagger c_{i,\sigma}
- \sum_{\la i,j \ra,\sigma} ( t_{i,j} - \alpha y_{i,j} )
( c_{i,\sigma}^\dagger c_{j,\sigma} + {\rm h.c.} )
+ \frac{K}{2} \sum_{\la i,j \ra} y_{i,j}^2, \\
H_{\rm int} &=& U \sum_{i} 
(c_{i,\uparrow}^\dagger c_{i,\uparrow} - \frac{n_{\rm el}}{2})
(c_{i,\downarrow}^\dagger c_{i,\downarrow} 
- \frac{n_{\rm el}}{2}) \nonumber \\
&+& \sum_{i,j} W(r_{i,j}) 
(\sum_\sigma c_{i,\sigma}^\dagger c_{i,\sigma} - n_{\rm el})
(\sum_\tau c_{j,\tau}^\dagger c_{j,\tau} - n_{\rm el}).
\eneqa
In Eq. (1), the first term $H_{\rm pol}$ is the tight 
binding model along the polymer backbone with electron-phonon 
interactions which couple electrons with modulation modes of 
the bond lengths, and the second term $H_{\rm int}$ is the 
Coulomb interaction potentials among electrons.  In Eq. (2), 
$E_i$ is the site energy at the $i$th site; $t_{i,j}$ is the 
hopping integral between the nearest neighbor $i$th and $j$th 
sites in the ideal system without bond alternations; $\alpha$ 
is the electron-phonon coupling constant that modulates 
the hopping integral linearly with respect to the bond 
variable $y_{i,j}$ which measures the magnitude of the 
bond alternation of the bond $\la i,j \ra$; $y_{i,j} > 0$ 
for longer bonds and $y_{i,j} < 0$ for shorter bonds (the 
average of $y_{i,j}$ is taken to be zero); $K$ is the 
harmonic spring constant for $y_{i,j}$; and the sum is 
taken over the pairs of neighboring atoms.  Equation (3) is 
the Coulomb interactions among electrons.  Here, $n_{\rm el}$ 
is the average number of electrons per site; $r_{i,j}$ is 
the distance between the $i$th and $j$th sites; and 
\beeq
W(r) = \frac{1}{\sqrt{(1/U)^2 + (r/a V)^2}}
\eneq
is the parametrized Ohno potential.  The quantity $W(0) = U$ 
is the strength of the onsite interaction; $V$ means the 
strength of the long-range part ($W(r) \sim aV/r$ in the limit 
$r \gg a$); and $a$ is the mean bond length.

Excitation wavefunctions of the electron-hole pair are 
calculated by the Hartree-Fock approximation followed 
by the single excitation configuration interaction method.  
This method, which is appropriate
for the cases of moderate Coulomb interactions -- strengths 
between negligible and strong Coulomb interactions -- is 
known as the intermediate exciton theory in the literatures [7,8].
We write the singlet electron-hole excitations as
\beeq
|\mu, \lambda \rangle = \frac{1}{\sqrt{2}}
(c_{\mu, \uparrow}^\dagger c_{\lambda, \uparrow} 
+ c_{\mu, \downarrow}^\dagger c_{\lambda, \downarrow} )
| g \rangle,
\eneq
where $\mu$ and $\lambda$ mean unoccupied and occupied states, 
respectively, and $| g \rangle$ is the Hartree-Fock ground state.  
The general expression of the $\kappa$th optical excitation is:
\beeq
| \kappa \rangle = \sum_{(\mu,\lambda)} D_{\kappa,(\mu,\lambda)}
| \mu, \lambda \rangle.
\eneq
After inserting the relation with the site representation 
$c_{\mu,\sigma} = \sum_i \alpha_{\mu,i} c_{i,\sigma}$, we obtain
\beeq
| \kappa \rangle = \frac{1}{\sqrt{2}} \sum_{(i,j)} 
B_{\kappa,(i,j)} (c_{i,\uparrow}^\dagger c_{j,\uparrow} 
+ c_{i,\downarrow}^\dagger c_{j,\downarrow} ) | g \rangle,
\eneq
where 
\beeq
B_{\kappa,(i,j)} = \sum_{(\mu,\lambda)} D_{\kappa,(\mu,\lambda)}
\alpha_{\mu,i}^* \alpha_{\lambda,j}.
\eneq
Thus, $|B_{\kappa,(i,j)}|^2$ is the probability that an electron 
locates at the $i$th site and a hole is at the $j$th site.

We shall define the following quantity:
\beeq
P_\kappa = \sum_{i \in M} \sum_{j \in M} |B_{\kappa,(i,j)}|^2,
\eneq
where $M$ is a set of sites within a single monomer, in other
words, a set of carbon sites included in the brackets of each
polymer shown in Fig. 1.  When $P_\kappa > 1/N_{\rm m}$ 
($N_{\rm m}$ is the number of monomers used in the calculation
of periodic polymer chains), the electron and hole favor to 
have large amplitudes in the same single monomer.  Then, this 
excited state is identified as a short-range exciton.  On the 
other hand, when $P_\kappa < 1/N_{\rm m}$, the excited state is 
characterized as a long-range exciton.  This characterization 
method is performed for all the photoexcited states 
$| \kappa \rangle$, and a long-range component in the optical 
absorption spectrum is extracted from the total absorption.

\section{PPP}
\subsection{Ring torsion effects}

In this section, the torsion angle between neighboring phenyl 
rings $\Psi$ is taken into account in the hopping integral 
without ring torsions as $t-\alpha y \Rightarrow (t-\alpha y) 
{\rm cos} \Psi$ in the model reviewed above.  Here, $y$ is the 
length change of the corresponding bond.  The parameter values 
used in this section are $\alpha = 2.59t$/\AA, 
$K=26.6t$/\AA$^2$, $U=2.5t$, and $V=1.3t$.  They have been 
determined by comparison with experiments of PPV [7], and have 
been used in [1,8].  Most of the quantities in the energy units 
are shown by the unit of $t$ in this paper.  The model is solved 
by the mean field approximation and electron-hole excitations 
are calculated by the intermediate exciton formalism.  The 
long-range component of photoexcited states is characterized 
as we have shown in the previous section.

Figure 2 shows the optical absorption spectra calculated
for the system with the torsion angle $\Psi = 23^\circ$ 
[5] and the monomer number $N_{\rm m}=20$.  The spectral 
shapes are nearly independent of the monomer number at 
$N_{\rm m}=20$, as reported in [3].  The polymer without
ring torsions with the open boundary is in the $x$-$z$ plane.  
The electric field of light is parallel to the chain and in 
the direction of the $x$-axis in Fig. 2 (a), and it is 
perpendicular to the chain and is along with the $y$-axis in 
Fig. 2 (b).  The electric field is along with the $z$-axis
in Fig. 2 (c).  The orientationally averaged spectra with 
respect to the electric field are shown in Fig. 2 (d).
In each figure, the bold line shows the total absorption
and the thin line shows the contribution from the long-range
excitons where photoexcited electron and hole are separated
over more than the spatial extent of the single monomer.
In Fig. 2 (a), there are two main features around 1.4$t$
and 2.4$t$, where quantities with energy dimensions are expressed 
in units of $t$.  There is a feature of long-range excitons 
around $\sim 2.0t$ at the higher energy side of the 1.4$t$ main 
peak.  In contrast, the 2.4$t$ feature does not have so strong 
long-range component due to the almost localized nature of 
excitons.  This is similar to the calculations in [3], and the 
ring torsions do not change the almost localized character of 
the feature around 2.4$t$ [9].  Figure 2 (b) shows the case 
where the electric field is parallel to the $y$-axis.  If the 
ring torsions are not present, the oscillator strengths are 
zero in this case.  However, they are finite and the maximum
of the bold line in Fig. 2 (b) is one order of magnitudes
smaller than that of Fig. 2 (a).  Two main features of the 
energies of 2.2$t$ and 2.8$t$ are derived from those of the 
case with the electric field in the $z$-direction, shown in 
Fig. 2 (c).  These features in Fig. 2 (c) contribute dominantly, 
and their oscillator strengths are of the same order as those 
of Fig. 2 (a).  There are long-range components of these 
two features around the energies about 2.5$t$ and 3.3$t$.  
Figure 2 (d) shows the optical spectra where the electric 
field is orientationally averaged.  The overall spectral 
shape is similar to that in the case without ring torsions 
$\Psi=0$ [3].  However, the spectral width decreases slightly
due to the smaller hopping interactions between neighboring
phenyl rings.

Next, we vary the torsion angle $\Psi$ in the Hamiltonian,
and look at changes in optical spectra.  We calculate the
long-range component in the total oscillator strengths, as
we have done in [1-3].  We search for its variations as a
function of $\Psi$.  Figure 3 shows the calculated results.
The closed squares are the data of the orientationally 
averaged absorption.  The open circles, squares, and triangles 
indicate the data for the cases with the electric field 
parallel to the $x$-, $y$-, and $z$-axes, respectively.
We find that the average long-range component weakly depends 
on $\Psi$, while $\Psi$ is as large as about 40$^\circ$.
Thus, we have examined ring torsion effects on optical 
excitations of PPP.  We have found that the magnitude of 
the torsion $\Psi \sim 23^\circ$ does not change the component 
of the long-range excitons so much from that of the torsion
free system.  However, the long-range component suddenly 
deceases at larger $\Psi$, and becomes zero at $\Psi=90^\circ$.  
This is a natural conclusion of the broken conjugations due to
strong ring torsions.  The long-range components with the
field in the $y$- and $z$-directions are generally larger
than that of the $x$-direction case.  This property has been
seen in Ref. [3], too.

\subsection{Torsional disorder}

Ring torsional disorder effects are taken into account by
the Gaussian distribution function of the standard deviation 
$\delta \Psi$.  This quantity is changed within $0^\circ \leq
\delta \Psi \leq 90^\circ$.  In this subsection, we report 
the results averaged over 1000 samples of Gaussian disorder 
for the system with $N_{\rm m} = 10$.  This $N_{\rm m}$ is half of that
of the previous subsection in order to take sample average
for distribution of disorder.

Figure 4 shows the sample-averaged long-range component (LRC) against 
the disorder strength $\delta \Psi$.  The static torsion angle, 
in other words, the torsion angle of the system without disorder 
is fixed as $\Psi = 23^\circ$.  The left axis is scaled by
the LRC of the orientationally averaged absorption in the system 
without disorder.  The closed squares are for the orientationally 
averaged absorption.  Therefore, this quantity is unity at $\delta \Psi
= 0$.  The open circles, squares, and triangles indicate 
the data for the cases with the electric field parallel to the 
$x$-, $y$-, and $z$-axes, respectively.  The data with electric
field parallel to the $y$-axis are smaller than the data 
orientationally averaged.  Those for cases with electric field 
parallel to the $x$- and $z$-axes are larger than the data 
shown by the closed squares.  The most important result is 
that the dependence of LRC on $\delta \Psi$ is very weak.  
The LRC does not change so much even if $\delta \Psi$ increases
as much as $90^\circ$.  This indicates that the long-range 
excitons survive against extra perturbations by the torsional 
disorder.

\section{PAN}
\subsection{Ring torsion effects}

It is of some interests to investigate ring torsion effects 
in conjugated polymers where torsion angles are larger than 
that of PPP.  In this section, we look at torsion effects 
in PAN as a typical example of such the polymers.  The polymer 
structure is shown schematically in Fig. 1 (b).  The zigzag
geometry of PAN is taken into account in the actual 
calculations, even though it is not shown in Fig. 1 (b).  For 
model parameters, we have assumed the negative site energy
$E_{\rm N} = -0.571t$ at the N sites ($E_{\rm C} = 0$ at the
C sites), the hopping integral $t_{\rm C-N} = 0.8t$ between 
carbon and nitrogen ($t_{\rm C-C} = t$ as in PPP), the onsite 
interaction strength $U=2t$, and the magnitude of long-range 
interactions $V=1t$.  The parameters of $E_{\rm N}$ and 
$t_{\rm C-N}$ are taken from [10] as representative values.
We show the results with the monomer number $N_{\rm m}=20$.
We have taken several combinations of Coulomb interaction 
parameters and have looked at changes of results.  We have
found that the the long-range component as a function of $\Psi$
depends weakly on $U$ and $V$.  Thus, we report the results for
one combination of Coulomb parameters.

Figure 5 show the calculated optical absorption spectra of PAN
with the torsion angle $\Psi = 56^\circ$.  The three dimensional
coordinates of the polymer are similar to those of PPP in Fig. 2.
In Fig. 5 (a), the electric field is along with the polymer axis,
and is in the $x$-direction.  There are two main features around
0.6$t$ and 2.6$t$.  The former has smaller long-range exciton
feature than the latter.  As we observe in the band structure
of PAN [10], the lowest optical excitation among the band gap 
has a certain magnitude of dispersions, and thus its long-range 
component might be observed.  However, this lowest exciton is
dipole forbidden in the system without ring torsions when the 
electric field is parallel to the chain direction.  There is 
the second lowest optical excitation, which is almost localized
and is dipole allowed in the system without torsions, at the excitation
energy near the lowest exciton.  Thus, the long-range component
of the 0.6$t$ feature is suppressed.  Figure 5 (b) shows the 
weak optical absorption originated from the finite torsions
as we have found for the PPP in Fig. 2 (b).  Figure 5 (c) shows
the optical spectra when the electric field is in the $z$-direction.
Two features around 1.2$t$ and 2.6$t$ have certain magnitudes
of oscillator strengths due to long-range excitons.  This 
qualitative  property is similar to that in Fig. 2 (c).
Then, Fig. 5 (d) shows the optical spectra where orientational
average is performed.  We find excitation features superposed
from those of Figs. 5 (a), (b), and (c).

Next, we show the long-range component of the oscillator
strength as a function of the torsion angle $\Psi$ in Fig. 6.
In contrast to PPP, the long-range component decreases smoothly
as $\Psi$ increases.  The long-range component of the total
oscillator strengths at the observed $\Psi = 56^\circ$ is about 
half of the magnitude of the system without ring torsions. 
Thus, the long-range excitons in PAN can be hindered by ring 
torsions more easily than those in PPP.  This is due to the fact 
that the torsion angle is larger in PAN, and also that two bonds, 
whose hopping integrals are modulated as $t {\rm cos}\Psi$,
are present between neighboring phenyl rings in PAN while one
such kind of bond is present between phenyls in PPP.

\subsection{Torsional disorder}

Ring torsional disorder effects are studied for PAN with using
the Gaussian distribution function of the standard deviation 
$\delta \Psi$ which is changed within $0^\circ \leq \delta \Psi 
\leq 90^\circ$.  The sample number of disorder is 1000, and
$N_{\rm m} = 10$ in this subsection.

Figure 7 shows the sample-averaged LRC as a function of the disorder 
strength $\delta \Psi$.  The torsion angle of the system without 
disorder is fixed as $\Psi = 56^\circ$.  The left axis is 
normalized as has been done for PPP in Fig. 4.  The closed 
squares are for the orientationally averaged absorption.  The 
open circles, squares, and triangles indicate the data for 
the cases with the electric field parallel to the $x$-, $y$-, 
and $z$-axes, respectively.  The four quantities at $\delta 
\Psi = 30^\circ$ decrease by the factor about half from the magnitudes at 
$\delta \Psi=0$.  Therefore, the LRC of PAN is hindered 
by the torsional disorder more easily than that of PPP.
Such the remarkable contrast between PPP and PAN is similar
to the contrast of the static torsion angle effects which
we have discussed in the previous subsection.  Thus, we 
have found that the spatial extent of excitons in PAN is 
easily hindered by the torsional disorder as well as by
the static ring torsions.

\section{Summary}

We have examined the ring torsion effects which might interrupt
delocalizations of excitons in the chain direction of the conjugated
polymers PPP and PAN.  Long-range excitons in the optical excitations
have been characterized, and the long-range component of oscillator
strengths has been calculated as a function of the torsion angle.
We have shown that the torsion effects in PPP are relatively small. 
In contrast, the torsions of PAN decrease the long-range component 
by about half of the magnitudes from that of the torsion-free system.
Next, ring torsional disorder effects simulated by the Gaussian 
distribution function have been analyzed.  The long-range component 
of the total oscillator strengths after sample average is nearly 
independent of the disorder strength $\delta \Psi$ in the PPP case, 
while that of the PAN decreases easily as $\delta \Psi$ becomes 
stronger.

\mbox{}

\begin{flushleft}
{\bf Acknowledgements}
\end{flushleft}

Useful discussion with Y. Shimoi, S. Abe, and K. Murata
is acknowledged.  Numerical calculations have been 
performed on the DEC AlphaServer of Research Information
Processing System Center (RIPS), Agency of Industrial 
Science and Technology (AIST), Japan.

\pagebreak
\begin{flushleft}
{\bf References}
\end{flushleft}

\noindent
$[1]$ K. Harigaya, J. Phys. Soc. Jpn. {\bf 66}, 
1272 (1997).\\
$[2]$ K. Harigaya, J. Phys.: Condens. Matter {\bf 9},
5253 (1997).\\
$[3]$ K. Harigaya, J. Phys.: Condens. Matter {\bf 9},
5989 (1997).\\
$[4]$ D. Chen, M. J. Winokur, M. A. Masse, and F. E. Karasz,
Phys. Rev. B {\bf 41}, 6759 (1990).\\
$[5]$ J. L. Baudour, Y. Delugeard, and P. Rivet,
Acta Crystallogr. B {\bf 34}, 625 (1978).\\
$[6]$ M. E. Jozefowicz, R. Laversanne, H. H. S. Javadi, 
A. J. Epstein, J. P. Pouget, X. Tang, and A. G. MacDiarmid,
Phys. Rev. B {\bf 39}, 12958 (1989).\\
$[7]$ Y. Shimoi and S. Abe, Synth. Met. {\bf 78}, 219 (1996).\\
$[8]$ K. Harigaya, Y. Shmoi, and S. Abe, in {\sl Proceedings
of the 2nd Asia Symposium on Condensed Matter Phytophysics}
(Nara, 1996), p. 25.\\
$[9]$ Z. G. Soos, S. Etemad, D. S. Galv\~{a}o, and S. Ramasesha,
Chem. Phys. Lett. {\bf 194}, 341 (1992).\\
$[10]$ J. M. Ginder and A. J. Epstein, Phys. Rev. B
{\bf 41}, 10674 (1990).\\

\pagebreak

\begin{flushleft}
{\bf Figure Captions}
\end{flushleft}

\mbox{}

\noindent
Fig. 1.  Polymer structures of (a) PPP and (b) PAN.
These figures are only schematic.  Geometries with ring 
torsions and the zigzag chain structure of PAN are used 
in the actual calculations.

\mbox{}

\noindent
Fig. 2. Optical absorption spectra of the PPP shown in arbitrary
units.  The polymer axis is parallel to the $x$-axis.  The polymer 
without ring torsions is in the $x$-$z$ plane.  The electric field 
of light is parallel to the chain and in the direction of the 
$x$-axis in (a), and it is perpendicular to the chain and is along 
with the $y$-axis in (b).  It is along with the $z$-axis
in (c).  The orientationally averaged spectra are shown in (d).
The number of the PPP monomer units is $N_{\rm m}= 20$.  
The bold line is for the total absorption.  The thin line 
indicates the absorption of the long-range component.  
The Lorentzian broadening $\gamma = 0.15 t$ is used.

\mbox{}

\noindent
Fig. 3.  Long-range component of the optical absorption
spectra as a function of the torsion angle in PPP.  The 
monomer unit number is $N_{\rm m}=20$.  The closed squares 
are for the orientationally averaged absorption.  The open 
circles, squares, and triangles indicate the data for the 
cases with the electric field parallel to the $x$-, $y$-,
and $z$-axes, respectively.

\mbox{}

\noindent
Fig. 4.  Long-range component (LRC) of the optical absorption
spectra as a function of the torsional disorder strength
$\delta \Psi$ in PPP with fixed $\Psi = 23^\circ$.  The monomer 
unit number is $N_{\rm m}=10$, and the number of disorder samples 
is 1000.  The closed squares are for the orientationally averaged 
absorption.  The open circles, squares, and triangles indicate 
the data for the cases with the electric field parallel to the 
$x$-, $y$-, and $z$-axes, respectively.  The left axis is 
normalized by the LRC of the orientationally averaged absorption 
in the system without disorder.

\mbox{}

\noindent
Fig. 5. Optical absorption spectra of the PAN shown in arbitrary
units.  The polymer axis is parallel to the $x$-axis.  The polymer 
without ring torsions is in the $x$-$z$ plane.  The electric field 
of light is parallel to the chain and in the direction of the 
$x$-axis in (a), and it is perpendicular to the chain and is along 
with the $y$-axis in (b).  It is along with the $z$-axis
in (c).  The orientationally averaged spectra are shown in (d).
The number of the PAN monomer units is $N_{\rm m}= 20$.  
The bold line is for the total absorption.  The thin line 
indicates the absorption of the long-range component.  
The Lorentzian broadening $\gamma = 0.15 t$ is used.

\mbox{}

\noindent
Fig. 6.  Long-range component of the optical absorption
spectra as a function of the torsion angle in PAN.  The 
monomer unit number is $N_{\rm m}=20$.  The closed squares 
are for the orientationally averaged absorption.  The open 
circles, squares, and triangles indicate the data for the 
cases with the electric field parallel to the $x$-, $y$-,
and $z$-axes, respectively.

\mbox{}

\noindent
Fig. 7.  Long-range component (LRC) of the optical absorption
spectra as a function of the torsional disorder strength
$\delta \Psi$ in PAN with fixed $\Psi = 56^\circ$.  The monomer 
unit number is $N_{\rm m}=10$, and the number of disorder samples 
is 1000.  The closed squares are for the orientationally averaged 
absorption.  The open circles, squares, and triangles indicate 
the data for the cases with the electric field parallel to the 
$x$-, $y$-, and $z$-axes, respectively.  The left axis is 
normalized by the LRC of the orientationally averaged absorption 
in the system without disorder.

\end{document}